\documentclass[trackchanges]{aastex63}

\usepackage{graphicx}
\usepackage{rotating}
\usepackage{hyphenat}
\usepackage{footnote}
\usepackage{url}
\usepackage{color,soul}
\usepackage{natbib}
\usepackage{amsmath}
\usepackage{wrapfig}

\usepackage{float}
\usepackage{url}
\usepackage{CJK}

\newcounter{supp}[section]

\submitjournal{PSJ}
\received{2023 October 16}
\revised{2024 January 16}
\accepted{2024 January 30}
\shorttitle{RhoPop}

\shortauthors{Schulze et al.}

\graphicspath{./}

\begin{document}
\begin{CJK*}{UTF8}{gbsn}

\title{A gap in the densities of small planets orbiting M dwarfs: rigorous statistical confirmation using the open-source code \texttt{RhoPop}}

\correspondingauthor{Joseph Schulze}
\email{schulze.61@osu.edu}

\author[0000-0003-3570-422X]{J.G. Schulze}
\affiliation{School of Earth Sciences, The Ohio State University, 125 South Oval Mall, Columbus OH, 43210, USA}

\author[0000-0002-4361-8885]{Ji Wang (王吉)}
\affiliation{Department of Astronomy, The Ohio State University, 140 West 18th Avenue, Columbus Ohio, 43210, USA}

\author[0000-0001-7258-1834]{J. A. Johnson}
\affiliation{Department of Astronomy, The Ohio State University, 140 West 18th Avenue, Columbus Ohio, 43210, USA}

\author[0000-0003-0395-9869]{B. S. Gaudi}
\affiliation{Department of Astronomy, The Ohio State University, 140 West 18th Avenue, Columbus Ohio, 43210, USA}

\author[0000-0003-1445-9923]{R. Rodriguez Martinez}
\affiliation{Center for Astrophysics \textbar \ Harvard \& Smithsonian, 60 Garden St, Cambridge, MA 02138, USA}

\author[0000-0001-8991-3110]{C.T. Unterborn}
\affiliation{Southwest Research Institute, 6220 Culebra Rd, San Antonio, TX 78238, USA}

\author[0000-0001-5753-2532]{W.R. Panero}
\affiliation{School of Earth Sciences, The Ohio State University, 125 South Oval Mall, Columbus OH, 43210, USA}
\affiliation{Division of Earth Sciences, U.S. National Science Foundation, Alexandria, VA, USA}


\begin{abstract}
Using mass-radius-composition models, small planets ($\mathrm{R}\lesssim 2 \mathrm{R_\oplus}$) are typically classified into three types: iron-rich, nominally Earth-like, and those with solid/liquid water and/or atmosphere. These classes are generally expected to be variations within a compositional continuum. Recently, however, \citet{Luque_2022} observed that potentially Earth-like planets around M dwarfs are separated from a lower-density population by a density gap. Meanwhile, the results of \citet{Adibekyan_2021} hint that iron-rich planets around FGK stars are also a distinct population. It therefore remains unclear whether small planets represent a continuum or multiple distinct populations. Differentiating the nature of these populations will help constrain potential formation mechanisms. We present the \texttt{RhoPop} software for identifying small-planet populations. \texttt{RhoPop} employs mixture models in a hierarchical framework and a nested sampler for parameter and evidence estimates. Using \texttt{RhoPop}, we confirm the two populations of \citet{Luque_2022} with $>4\sigma$ significance. The intrinsic scatter in the Earth-like subpopulation is roughly half that expected based on stellar abundance variations in local FGK stars, perhaps implying M dwarfs have a smaller spread in the major rock-building elements (Fe, Mg, Si) than FGK stars. We apply \texttt{RhoPop} to the \citet{Adibekyan_2021} sample and find no evidence of more than one population. We estimate the sample size required to resolve a population of planets with Mercury-like compositions from those with Earth-like compositions for various mass-radius precisions. Only 16 planets are needed when $\sigma_{M_p} = 5\%$ and $\sigma_{R_p} = 1\%$. At $\sigma_{M_p} = 10\%$ and $\sigma_{R_p} = 2.5\%$, however, over 154 planets are needed, an order of magnitude increase.
\end{abstract}

\section{Introduction}
\label{sect:intro}
Small planets ($\mathrm{R}\lesssim 2 \mathrm{R_\oplus}$) are among the most common in the Galaxy. Standard formation models predict that the relative amounts of the major rock-building elements (Fe/Mg and Si/Mg) in small planets should reflect that of their host stars \citep[e.g.,][]{Bond_2010a, Bond_2010b, Elser_2012, Johnson_2012, Thiabaud_2015}. The inferred compositions of observed small exoplanets \citep{Schulze21, Adibekyan_2021, Unterborn_2022} and rocky exoplanetary material \citep{Bonsor_2021} are consistent with this host-planet compositional connection at current observational precisions except in extreme cases, e.g., Kepler-107 c \citep{Bonomo19_Kep107_Nature} and HD 137496 b \citep{Silva_2022}. In the Solar System, Earth and Mars have compositions that are consistent with the solar abundances of the major rock-building elements \citep[e.g.,][]{Wanke94_Chem_and_Acc_Hist_Mars, Lodders03, Wang19_ElementalAbund_Vol_Trend, Yoshizaki_McDonough20}. The composition of Venus is unknown but expected to be consistent with Earth \citep[e.g.,][]{Zharkov83_IntStruct_Venus, Aitta_2012, Dumoulin_2017}. Mercury, in contrast, has a 200-400\% overabundance of Fe relative to the Sun, indicative of a fundamentally distinct formation regime and/or subsequent evolution from the other small planets in the Solar System \citep[e.g.,][]{ CAMERON_1985, BENZ_1988, Wurm_2013, Anders_nucleation}. If we, therefore, assume that small planets are barren and rocky, then inconsistencies between inferred planet composition and host composition indicate deviations from standard planet formation models/theories.

Recently, \citet{Luque_2022} observed a bimodal density distribution for 27 small planets in orbit around M dwarf hosts. The authors interpret this bimodality as two compositionally distinct subpopulations, one with Earth-like compositions and one with 50\% water and 50\% rock by mass, but do not provide robust statistical evidence to support this claim or a comparison with host abundances of the major rock-building elements. There is a paucity of known M dwarf abundances of the major rock-building elements as measuring M dwarf abundances is a non-trivial problem and an active area of research \citep[e.g.,][]{Allard2000, Souto22}. As such, it is unclear whether there are indeed two subpopulations of planets around M dwarfs and to what degree these planets deviate from the host-planet compositional connection. \citet{Adibekyan_2021} show that 22 likely rocky small planets in orbit around FGK stars have compositions correlated with their hosts' but not with a 1:1 relationship. The authors identify a class of 5 planets with a Mercury-like overabundance of iron separated from the general star-planet compositional link by a visually apparent, but not statistically conclusive, gap in density. \citet{Barros_2022} identify two additional planets around HD 23472 that appear to belong to the same iron-enriched population identified in \citet{Adibekyan_2021}. Together these studies suggest that the small-planet density distribution may not be a continuum. Instead, planet formation mechanisms might create up to three distinct classes of small planets: (1) those that follow the star-planet compositional link, (2) those enriched in water/atmosphere and/or depleted in iron, and (3) those enriched in iron and/or depleted in silicates. 

Statistical host-planet composition comparisons like that of \citet{Adibekyan_2021} provide a powerful tool for confirming iron-enriched/silicate-depleted or water-enriched/iron-depleted small planets \citep[e.g.,][]{Santerne18, Schulze21, Romy_K2106_paper, Silva_2022, Unterborn_2022}. There is, however, a paucity of planets with known densities \textit{and} host abundances of Fe, Mg, and Si, and no single catalog lists all of these quantities simultaneously. Combining the NASA Exoplanet Archive\footnote{\href{https://exoplanetarchive.ipac.caltech.edu/index.html}{https://exoplanetarchive.ipac.caltech.edu/index.html}} with the Hypatia Catalog stellar abundance database\footnote{\href{https://www.hypatiacatalog.com/hypatia}{https://www.hypatiacatalog.com/hypatia}} \citep{Hinkel_2014}, we estimate there are approximately 43 planets with measured densities and host Fe, Mg, and Si abundances. Of these, only 12 are small planets. Direct host-planet comparisons are therefore limited to single planets or small samples \citep[e.g., 11, 22, and 7 in ][respectively]{Schulze21, Adibekyan_2021, Unterborn_2022}.

To overcome this lack of host-star abundances, some studies instead compare inferred small-planet compositions with a large sample of stellar abundances of the major rock-building elements \citep{Scora_2020, Plotnykov20, Unterborn_2022}. \citet{Unterborn_2022} used the Hypatia Catalog stellar abundance database \citep{Hinkel_2014} to quantify a nominally rocky planet zone (NRPZ): the expected 99.7\% range of rocky planet densities owing to the intrinsic spread in stellar compositions. Where the density and corresponding uncertainty of a small planet exclude it from the NRPZ, its bulk composition likely deviates from that predicted through standard formation models. Excluded planets whose densities exceed that of the NRPZ are then interpreted as having an iron enrichment and/or silicate depletion. Excluded planets with densities less than the NRPZ are interpreted as being iron-depleted, water-rich, or volatile-rich. Applying this technique to a large sample of small planets, \citet{Unterborn_2022} find 21 planets that are excluded from the NRPZ with $\geq 90\%$ confidence: 8 are likely iron-enriched and 13 are iron-depleted, water-rich, or volatile-rich.

It remains an open question whether the densities of small planets reflect a continuous variation in composition or multiple, distinct subpopulations. Robust identification of these subpopulations, or lack thereof, is an imperative first step to better constraining their formation and subsequent evolution. Direct host-planet compositional comparisons are a powerful method for identifying planets that deviate from standard formation models but are sample-size-limited, making it difficult to resolve distinct subpopulations of small planets. This limitation is often circumvented by instead comparing planet compositions with a large sample of stellar abundances. While the latter has been used to identify a number of individual planets that are inconsistent with the NRPZ, no study has investigated whether these planets form a continuum with the NRPZ or are members of fundamentally different subpopulations.

%\footnote{\href{https://anonymous.4open.science/r/RhoPop-7B86/}{https://anonymous.4open.science/r/RhoPop-7B86/}}

In this work, we present the open-source \texttt{RhoPop} \footnote{\href{https://zenodo.org/records/10368148}{https://zenodo.org/records/10368148}} code for robust identification of compositionally distinct populations of small planets. \texttt{RhoPop} models the density distribution of an input planet sample as a Gaussian mixture of up to three compositionally-distinct subpopulations in a hierarchical Bayesian framework. Similar models have been used for exoplanet population analyses, but small planets are treated as a single population \citep{Chen_2017_MR_forecasting} or assume a fixed composition for rocky materials \citep{Neil_2020, Neil_2022}. \texttt{RhoPop} uses compositional grids that span from pure-iron to pure-water planets that are normalized relative to the average composition of the NRPZ, so that model results are easily comparable with the expected compositional range of small water-/volatile-free planets. We describe the inner workings of our software in Sect. \ref{sect:methods}.  We use \texttt{RhoPop} to validate the results of \citet{Luque_2022} and \citet{Adibekyan_2021} in Sect. \ref{sect:Luque_results} and \ref{sect:adib_results}, respectively. In Sect. \ref{sect:n_planet_estimates}, we use \texttt{RhoPop} to estimate the minimum number of planets needed to identify a population of Mercury-like planets from a population of Earth-like planets for various mass and radius uncertainties.

\section{Methods}
\label{sect:methods}

\subsection{Density-mass-composition grids}
\label{Sect:compgrids}

We use the open-source \texttt{ExoPlex}\footnote{\href{https://github.com/CaymanUnterborn/ExoPlex}{https://github.com/CaymanUnterborn/ExoPlex}} mass-radius-composition calculator \citep{Unterborn_2022} to build density-mass grids from pure water worlds to pure-iron planets. ExoPlex finds the radius for a user-input planet mass and bulk composition by self-consistently solving the five coupled differential equations: the mass within a sphere, hydrostatic equilibrium, adiabatic temperature profile, Gauss's law of gravity in one dimension, and the thermally dependent equation of state. We assume a pure-Fe core and adopt ExoPlex's default equation of state (EOS) for liquid iron \citep{Anderson_94_liquidFe}. ExoPlex couples the thermodynamic database of \citet{stixrude_2011} with the \texttt{Perple\_X} thermodynamic equilibrium software \citep{perplex_connolly} to find the equilibrium mineralogy and corresponding material density throughout the mantle. For outer water layers, ExoPlex adopts the \texttt{Seafreeze} \citep{Journaux_2020} software for liquid water, Ice Ih, II, III, V, or VI. For Ice VI, ExoPlex couples the isothermal EOS of \citet{Journaux_2020} with the empirical equations of \citet{ASAHARA_2010} and \citet{Fei_1993}.

We build isocomposition rock and liquid/solid water grids over a mass range of 0.05 to 10 $M_\oplus$ in increments of $\Delta \log_{10} (\mathrm{M / M_\oplus}) \simeq 0.023$. For all planets, we assume the silicate mantle is Fe-free and fix the mantle molar ratios to Si/Mg = 0.79, Al/Mg = 0.07, and Ca/Mg = 0.09 per \citet{Unterborn_2022} as the Fe/Mg ratio is the primary control on density for water-free rocky planets \citep{Dorn_2015, Unterborn_2016}. For pure-rock compositions, we vary $\log_{10}(\mathrm{Fe/Mg})$ from -0.2 to 1.5 (Fe/Mg = 0.01-30) in increments of $\Delta \log_{10} (\mathrm{Fe/Mg}) \simeq 0.035$. Our range of Fe/Mg corresponds to a core mass fraction (CMF) range of 0.006 to 0.95. For water-rich compositions, we build a grid of WMF = 0.001, 0.005, 0.01, 0.025, and then in increments of $\Delta \mathrm{WMF} = 0.025$ from 0.025 to 1.0. We fix the interior rocky portion of these planets to have Fe/Mg = 0.71 corresponding to the average Hypatia values per \citet{Unterborn_2022}. For compositions between our computed grid lines, we linearly interpolate between the two nearest compositional neighbors to approximate the intermediate isocomposition line. 

We note that there is an overlap between our rock grid and water grid between WMF = 0 and 0.1-0.15, depending on mass, and Fe/Mg = 0 - 0.71 (CMF = 0-0.29), meaning there are two grid solutions for a given mass and radius in this regime: one with water and one without. This is a result of a well-known degeneracy when inferring planet composition from density alone \citep[e.g.,][]{Rogers_2015, Dorn_2015, Unterborn_2016} which persists despite our simplifications of a pure-Fe core and Fe-free silicate mantle. We circumvent this degeneracy by using a reduced rock grid from Fe/Mg = 0.71-30 (CMF = 0.29-0.95). For planet densities lower than a water-free planet with Fe/Mg = 0.71, \texttt{RhoPop} switches to the water grid. Where our models return a WMF = 0.0-0.15, however, we also report the corresponding best fit Fe/Mg and CMF for a water-free planet.

\begin{figure}[ht]
    \centering
    \includegraphics[width = 0.8\textwidth]{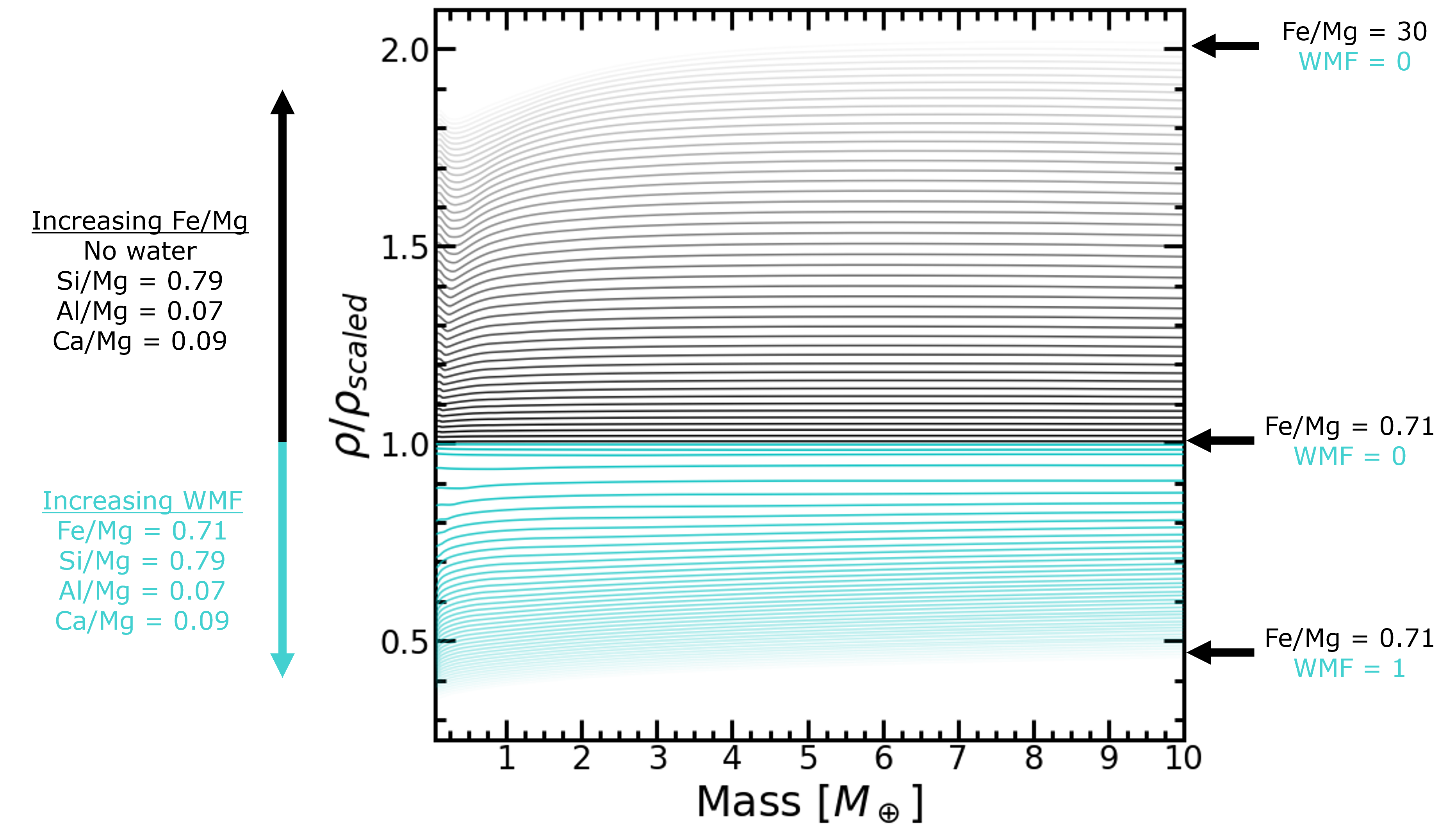}
    \caption{Water grid (cyan) and reduced rock grid (black). The $\rho_{scaled}$ parameter corresponds to the density at mass of a water- and volatile-free rocky planet with a liquid pure-iron core and Fe-free silicate mantle with Fe/Mg = 0.71, Si/Mg = 0.79, Ca/Mg = 0.09, and Al/Mg = 0.07.}
    \label{fig:comp_grid}
\end{figure}

Following the work of \citet{Luque_2022}, we normalize all planet densities to a scaled density parameter denoted as $\rho_{scaled}$. The scaled density parameter is a function of planet mass and a reference composition. It gives the density a planet of a given mass would have with the reference composition. Both our work and \citet{Luque_2022} assume volatile-/water-free compositions for reference. Where \citet{Luque_2022} uses a nominally `Earth-like' CMF = 0.325, however, we assume the average CMF = 0.29 of the Hypatia Catalog corresponding to Fe/Mg = 0.71, Si/Mg = 0.79, Ca/Mg = 0.09, and Al/Mg = 0.07. That is, $\rho_{scaled} (M) = \rho(M, \mathrm{CMF = 0.29})$. We choose this value so any planet populations inferred with \texttt{RhoPop} may be directly compared with the nominally rocky planet zone. We can then define a general density ratio parameter $\rho_{ratio}\equiv \rho/\rho_{scaled}$. For the ExoPlex-derived density-mass-grids, this is a function of mass and composition

\begin{equation}
    \rho_{ratio} = \rho(M, \mu) / \rho_{scaled}(M)
\end{equation}

\vspace{0.25 cm}
\noindent where $\mu$ is the composition in terms of a CMF from 0.29 to 1.0 or a WMF from 0 to 1.0. Our water grid and reduced rock grid are shown relative to $\rho_{scaled}$ in Fig. \ref{fig:comp_grid}.

\subsection{Hierarchical Bayesian Model Framework}
\label{Sect:HBMF} 
Inspired by the population insights of \citet{Chen_2017_MR_forecasting} and \citet{Neil_2022}, \texttt{RhoPop} employs Gaussian mixture models in a hierarchical Bayesian framework. Hierarchical Bayesian models (HBM) employ two sets of parameters, local and hyper. We treat each observed data point as a Gaussian random variable centered on its true value with a standard deviation equal to the observational uncertainty. Since the true values can not be known a priori, these are treated as free model parameters called `local parameters.' The parameters that govern the model that predicts the local parameters are the `hyperparameters'. Here the hyperparameters are those that describe the underlying populations of planets.

The local parameters of our HBM are the true planet masses, $M_{t}$, and density ratios, $\rho_{ratio, t}$. Observed planet mass, $M_\mathrm{ob}$, is modeled as a random Gaussian variable centered on the true mass, $M_t$, with a standard deviation equal to the observed mass uncertainty, $\sigma_{M_\mathrm{ob}}$. That is,

\begin{equation}
    M_\mathrm{ob}^{i} \sim \mathcal{N}\left(M_t^{i}, \sigma_{M_\mathrm{ob}^{i}}\right)
    \label{equ:m_observed}
\end{equation}

\noindent where the superscript $i$ denotes the $i$-th planet in the sample. For planets with asymmetric observational mass uncertainties, we take the average of the upper and lower uncertainties to calculate $\sigma_{M_\mathrm{ob}^{i}}$.  Evaluating the right-hand side of Eq. \ref{equ:m_observed} at $M_\mathrm{ob}^{i}$ gives the \textit{likelihood function} for $M_t^{i}$, which we denote as $\mathcal{L}_{m}^{i}$, i.e.,

\begin{equation}
    \mathcal{L}_{m}^{i} = \mathcal{N}\left(M_\mathrm{ob}^{i}\bigg|M_t^{i}, \sigma_{M_\mathrm{ob}^{i}}\right).
    \label{equ:like_mass}
\end{equation}

We link the local parameters to the hyperparameters that describe the planet populations using a Gaussian mixture model (GMM). Gaussian mixture models represent some global population as a combination of $N_c$ normally distributed subpopulations. Each subpopulation has a mixing weight, $w$, which is its fractional size relative to the global population. The sum of the mixing weights over all $N_c$ components must sum to unity, i.e.,

\begin{equation}
    \sum_{j=1}^{N_c} w^{j} = 1
    \label{equ:sum_mw}
\end{equation}

\noindent
where we use the superscript $j$ to denote the $j$-th subpopulation. We assume that each $j$ component of our Gaussian mixture models is centered on composition $\mu_{comp}^{j}$ with an intrinsic scatter of density ratios $\sigma_{\rho_{ratio, pop}^{j}}$. Within a given population, $\sigma_{\rho_{ratio, pop}^{j}}$ represents the variation that \textit{cannot} be accounted for by the observational uncertainties of the planets that belong to it. We then model the local parameter $\rho_{ratio, t}^{i}$ as a Gaussian mixture of all $N_c$ components

\begin{equation}
    \rho_{ratio, t}^{i} \sim \sum_{j=1}^{N_c}   w^{j} \times \mathcal{N}\left(f(M_t^{i}, \mu_{comp}^{j}), \sigma_{\rho_{ratio, pop}^{j}}\right).
    \label{equ:rho_ratio_true}
\end{equation}

The argument of the Gaussian in Eq. \ref{equ:rho_ratio_true}, $f(M_t^{i}, \mu_{comp}^{j})$, is calculated directly from the compositional grids (Sect. \ref{Sect:compgrids}) and represents the predicted density ratio at $M_{t}^{i}$ given a central composition of population $j$, $\mu_{comp}^{j}$. We then model the observed density ratio of the $i$-th planet, $\rho_{ratio, \mathrm{ob}}^{i}$, as being centered on $\rho_{ratio, t}^{i}$ with an error term corresponding to its observed uncertainty,

\begin{equation}
\begin{split}
    \rho_{ratio, \mathrm{ob}}^{i} \sim &\sum_{j=1}^{N_c}  \left[ w^{j} \times \mathcal{N}\left(f\left(M_t^{i}, \mu_{comp}^{j}\right), \sigma_{\rho_{ratio, pop}^{j}}\right) + \mathcal{N}\left(0, \sigma_{\rho_{ratio, \mathrm{ob}}^{i}}\right) \right] \\
    &= \sum_{j=1}^{N_c}  w^{j} \times \mathcal{N}\left(f\left(M_t^{i}, \mu_{comp}^{j}\right), \sqrt{\left(\sigma_{\rho_{ratio, pop}^{j}}\right)^2 + \left(\sigma_{\rho_{ratio,\mathrm{ob}}^{i}}\right)^2}\right).
    \label{equ:rho_ratio_obs}
    \end{split}
\end{equation}

\noindent
As with Eq. \ref{equ:m_observed}, evaluating the right-hand side of Eq. \ref{equ:rho_ratio_obs} at $\rho_{ratio, \mathrm{ob}}^{i}$ gives the likelihood function for $\rho_{ratio, t}^{i}$,

\begin{equation}
    \mathcal{L}_\rho^{i} = \sum_{j=1}^{N_c}  w^{j} \times \mathcal{N}\left(\rho_{ratio, \mathrm{ob}}^{i} \bigg| f\left(M_t^{i}, \mu_{comp}^{j}\right), \sqrt{\left(\sigma_{\rho_{ratio, pop}^{j}}\right)^2 + \left(\sigma_{\rho_{ratio,\mathrm{ob}}^{i}}\right)^2}\right).
    \label{equ:like_rho}
\end{equation}

\noindent
Finally, combing Equations \ref{equ:like_mass} and \ref{equ:like_rho}, taking the natural logarithm, and summing over all $N_p$ planets gives the log-likelihood function for the entire model as a function of the hyperparameters,

\begin{equation}
\begin{split}
    \ln \mathcal{L} &= \sum_{i = 1}^{N_p} \left[ \ln \mathcal{L}_{\rho}^{i} + \ln \mathcal{L}_{m}^{i} \right] \\
    &= \sum_{i = 1}^{N_p}\bigg[ \ln \left(  \sum_{j=1}^{N_c}  w^{j} \times \mathcal{N}\left(\rho_{ratio, \mathrm{ob}}^{i} \bigg| f\left(M_t^{i}, \mu_{comp}^{j}\right), \sqrt{\left(\sigma_{\rho_{ratio, pop}^{j}}\right)^2 + \left(\sigma_{\rho_{ratio,\mathrm{ob}}^{i}}\right)^2}\right) \right) \\
    & + \ln \left(\mathcal{N}\left(M_\mathrm{ob}^{i}\bigg|M_t^{i}, \sigma_{M_\mathrm{ob}^{i}}\right) \right) \bigg].
\end{split}
\label{equ:log_like}
\end{equation}

\begin{figure}[ht]
    \centering
    \includegraphics[width = 0.8\textwidth]{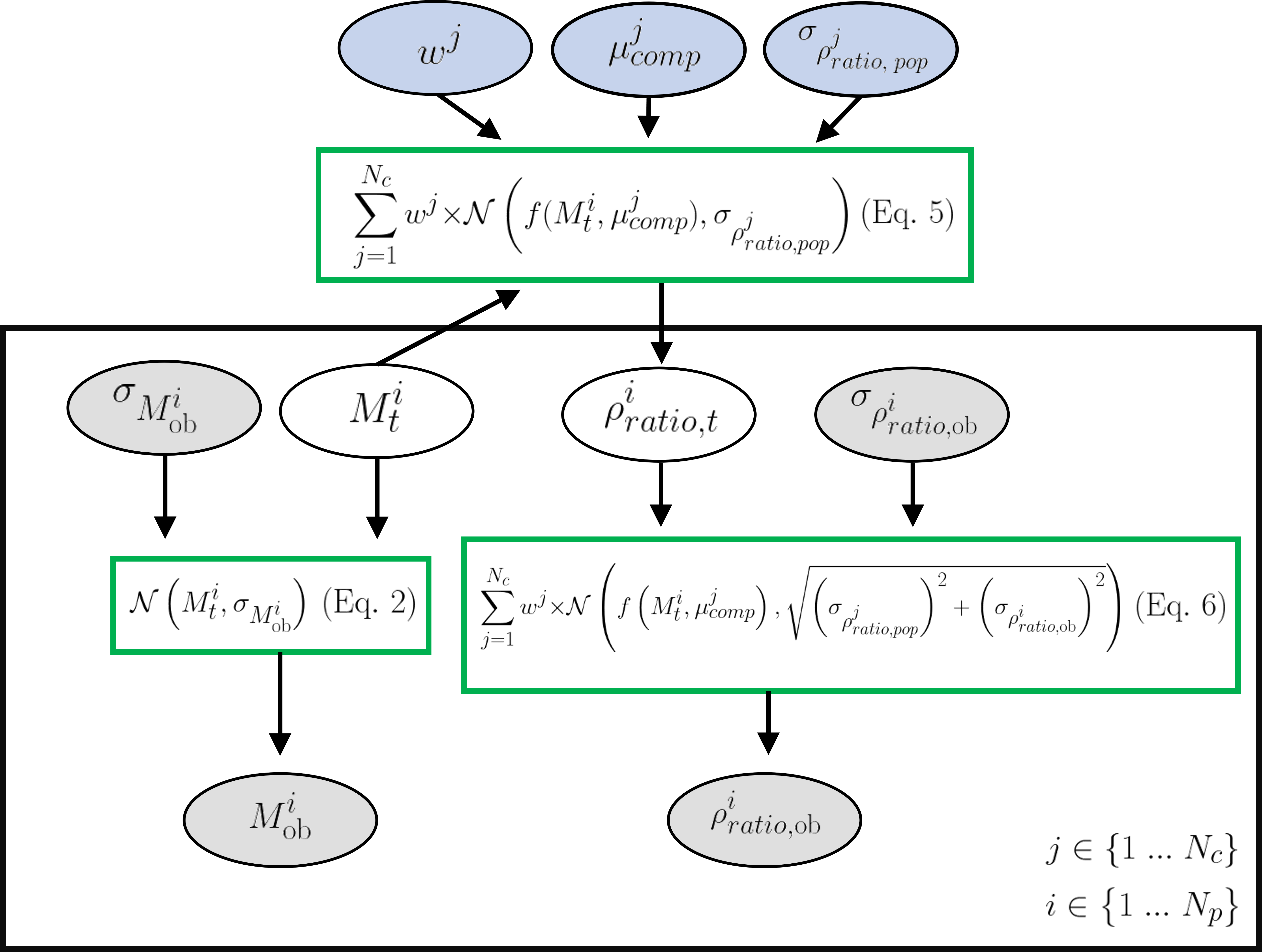}
    \caption{HBM framework. Hyperparameters are shown in blue, local parameters in white, and observed parameters in gray. The observed and local parameters always have $N_p$ members corresponding to the number of planets in the sample. $N_c = 1, 2,$ or 3 when the 1-, 2-, and 3-population model is considered, respectively. Observed masses and density ratios are modeled as Gaussian random variables centered on their unknown true values (local parameters) with a standard deviation equal to their observational uncertainties (Eq. \ref{equ:m_observed} and \ref{equ:rho_ratio_obs}, respectively). The local and hyperparameters are linked through the mixture model defined in Eq. \ref{equ:rho_ratio_true}.}
    \label{fig:hbm}
\end{figure}

\noindent
Our HBM framework is summarized schematically in Figure \ref{fig:hbm} and all parameters are summarized in Table \ref{tab:parameter_summary}. In this work, we compare models through Bayesian model evidence $\mathcal{Z}$: the integral of the likelihood (Eq. \ref{equ:log_like}) over the entire parameter prior space. Mathematically,

\begin{equation}
    P(\mathbf{D}|M) \equiv \mathcal{Z} = \int_{\Omega_\mathbf{\Theta}} \mathcal{L}(\mathbf{D} | \mathbf{\Theta}, \mathcal{M}) Pr(\mathbf{\Theta} | \mathcal{M}) d\mathbf{\Theta},
    \label{equ:evidence}
\end{equation}

\noindent
where $\mathbf{\Theta}$ represents a generic set of local parameters and hyperparameters, $Pr$ are the corresponding priors, and $\mathcal{M}$ is the model. We choose wide uninformative uniform priors on all hyperparameters to remain agnostic when validating previous results in Sections \ref{sect:Luque_results} and \ref{sect:adib_results}. We summarize all hyperparameter priors in Table \ref{tab:hp_priors}. \texttt{RhoPop} handles up to three populations, i.e., $N_c = 1$, 2, or 3. \texttt{RhoPop} makes no a priori assumptions about the central compositions of the populations $\mu_{comp}^{j}$ other than $\mu^1_{comp} \geq \mu^2_{comp}$ for $N_c = 2$ and $\mu^1_{comp} \geq \mu^2_{comp} \geq \mu^3_{comp}$ for $N_c = 3$. In the 3-population scenario, we allow $w^1$ and $w^2$ to vary from 0 to 1. However, if the sum of these 2 is greater than one, the likelihood function returns a $-\inf$, and that set of parameters is rejected since Eq. \ref{equ:sum_mw} must be satisfied. We assume priors of $M_t^{i} \sim \mathcal{N}(M_\mathrm{ob}^{i}, \sigma_{M_\mathrm{ob}^{i}})$ for all $i$ in $N_p$.

\begin{table}[ht]
\large
    \centering
    \begin{tabular}{|c|c|c|}
    \hline
         Parameter & Parameter Type &  Description \\ 
         \hline
         $\mu_{comp}^{j}$ & hyper & Central composition of population $j$ in WMF or CMF\\
         $\sigma_{\rho_{ratio, pop}^{j}}$ & hyper & intrinsic 1$\sigma$ density ratio scatter of population $j$ \\
         $w^{j}$ & hyper & Mixing weight of population $j$\\
         \hline
         $\rho_{ratio, t}^{i}$ & local & True density ratio of planet $i$ \\
         $M_t^{i}$ & local & True mass of planet $i$\\
         \hline
         $M_\mathrm{ob}^{i}$ & data & Central observed mass of planet $i$\\
         $\sigma_{M_\mathrm{ob}^{i}}$ & data & $1\sigma$ uncertainty in $M_\mathrm{ob}^{i}$\\
         $\rho_{ratio, \mathrm{ob}}^{i}$ & data & Central observed density ratio of planet $i$\\
         $\sigma \rho_{ratio, \mathrm{ob}}^{i}$ & data & $1\sigma$ uncertainty in $\rho_{ratio, \mathrm{ob}}^{i}$\\
         \hline

    \end{tabular}
    \caption{Summary of parameter definitions.}
    \label{tab:parameter_summary}
\end{table}

In practice, calculating $\mathcal{Z}$ analytically is not tractable and it must be computed numerically. As such, \texttt{RhoPop} uses the \texttt{dynesty}\footnote{\href{https://github.com/joshspeagle/dynesty}{https://github.com/joshspeagle/dynesty}} dynamic nested sampling software package with the default weighting scheme which is optimized to simultaneously estimate the parameter posteriors and model evidence. We point the reader to \citet{speagle20_dynesty} for more details.

\subsubsection{Model selection}
We adopt the methods outlined in \citet{Benneke_2013} which use Bayes factors to compare models of increasing complexity. A general Bayes factor $B_{ij}$ is defined as the ratio of the Bayesian evidence of model $i$ to the Bayesian evidence of model $j$, i.e., $B_{ij} \equiv \mathcal{Z}_{i} / \mathcal{Z}_{j}$. In terms of hypothesis testing, the denominator represents the evidence for the null hypothesis and the numerator is the alternative hypothesis. A value of $B_{ij} < 1$ suggests the data favor model $j$ over model $i$ i.e., the data are consistent with the null hypothesis. A $B_{ij}$ from 1-3 corresponds to inconclusive support for the more complex model $i$, and the null hypothesis cannot be rejected. A $B_{ij}$ from 3-12, 12-150, and $> 150$ correspond to weak, moderate, and strong support, respectively, for the alternative hypothesis. For a given sample in this work, we first run the 1- and 2-population models and compare via $B_{21} = \mathcal{Z}_2 / \mathcal{Z}_1$ where $\mathcal{Z}_1$ and $\mathcal{Z}_2$ correspond to the evidence for the 1- and 2-population models, respectively. Where there is significant support for the 2-population model ($B_{21} > 150$), we then run the 3-population model and compare it with the 2-population model via $B_{32}$. \citet{Benneke_2013}, and references therein, provide formulae to convert Bayes factors to the frequentist $p$-value (see their Eq. 10) and corresponding sigma significance $n_\sigma$ (see their Eq. 11). For all Bayes factors in this work, we use these formulae to calculate and report the equivalent $p$-values and $n_\sigma$ significance levels for the frequentist reader.

\begin{table}[ht]
\centering
\large
\begin{tabular}{l|l|l|l|}
\hline
\multicolumn{1}{|l|}{Param}                               & \multicolumn{1}{l|}{$N_c = 1$} & \multicolumn{1}{l|}{$N_c = 2$} & \multicolumn{1}{l|}{$N_c = 3$} \\ \hline
\multicolumn{1}{|l|}{$\mu_{comp}^1$}               & $\mathcal{U}$[1.0 WMF, 0.95 CMF] &   $\mathcal{U}$[1.0 WMF, 0.95 CMF] & $\mathcal{U}$[1.0 WMF, 0.95 CMF] \\
\multicolumn{1}{|l|}{$\sigma_{\rho_{ratio, pop}^1}$} &   $\mathcal{U}$[0,2]   &    $\mathcal{U}$[0,2]   &  $\mathcal{U}$[0,2]        \\
\multicolumn{1}{|l|}{$w^1$}                      &   1             &  $\mathcal{U}$[0,1] &  $\mathcal{U}$[0,1]     \\
\multicolumn{1}{|l|}{$\mu_{comp}^2$}               &                &  $\mathcal{U}[0,1]\times\mu_{comp}^1$             &  $\mathcal{U}[0,1]\times\mu_{comp}^1$     \\
\multicolumn{1}{|l|}{$\sigma_{\rho_{ratio, pop}^2}$} &                &                $\mathcal{U}$[0,2] &       $\mathcal{U}$[0,2]               \\
\multicolumn{1}{|l|}{$w^2$}                      &                &               1-$w^1$    &     $\mathcal{U}$[0,1]        \\
\multicolumn{1}{|l|}{$\mu_{comp}^3$}               &                &               &    $\mathcal{U}[0,1]\times\mu_{comp}^2$           \\
\multicolumn{1}{|l|}{$\sigma_{\rho_{ratio, pop}^3}$} &                &               &  $\mathcal{U}$[0,2]                  \\ 
\multicolumn{1}{|l|}{$w^3$}                      &                &               & 1$-w^1 - w^2$             \\ \hline

\end{tabular}
\caption{Hyperparameter priors. The default prior on $\mu_{comp}^1$ spans the entirety of our water and reduced rock grids.}
\label{tab:hp_priors}
\end{table}

\section{Validating Previous Results}

\subsection{\citet{Luque_2022} sample}
\label{sect:Luque_results}
The full sample of \citet{Luque_2022} contains 34 planets with mass and radius precisions of better than 25\% and 8\%, respectively, and radii less than 4 $R_\oplus$. The authors identify 7 planets as mini-Neptunes. We do not consider these planets further as they require an extended H/He envelope to explain and do not meet our definition of a small planet. We instead focus on the remaining 27 planets from which \citet{Luque_2022} observe a bimodal density distribution: a higher density population with 21 members and a lower density population with 6 members. The authors interpret the higher-density population as being purely rocky in composition and the lower-density population as having a rocky interior surrounded by a 50\% H$_2$O layer by mass.

\begin{figure}[t]
    \centering
    \includegraphics[width = 0.99\textwidth]{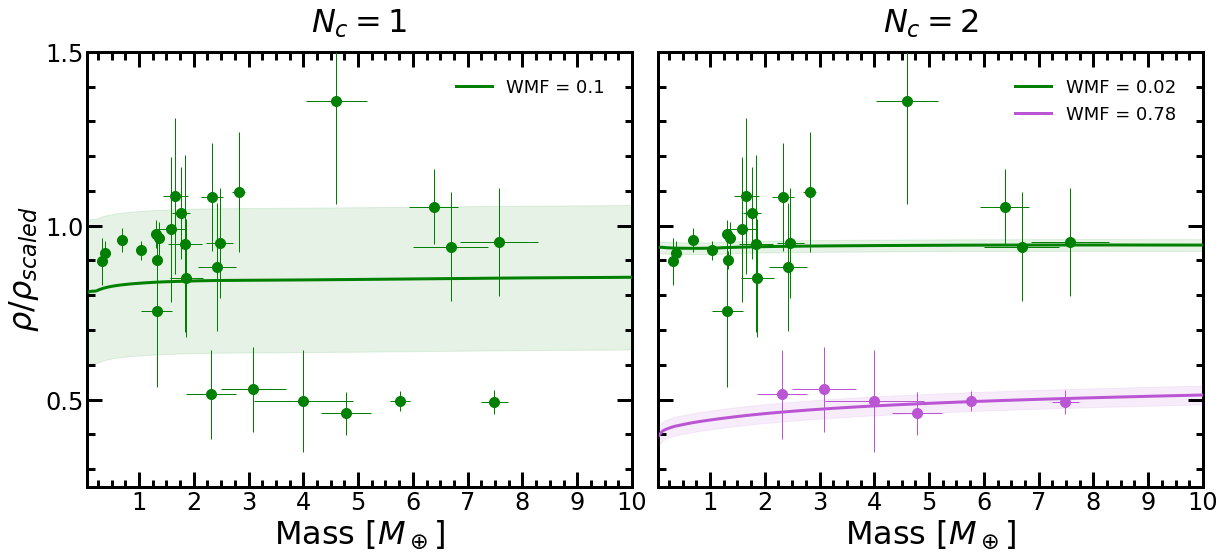}
    \caption{Density ratio as a function of mass for our $N_c = 1$ (left) and $N_c = 2$ (right) models applied to the \citet{Luque_2022} sample of small transiting planets in orbit around M-dwarfs. We color code the highest-density population as green and the lowest-density population as purple. The planets are assigned the same color as the population to which they most likely belong. The solid lines are the isocomposition lines for each $\mu_{comp}^{j}$. The bands are the corresponding intrinsic density ratio scatter parameter $\sigma_{\rho_{ratio,pop}}^{j}$. For all $\mu_{comp}^{j}$ and $\sigma_{\rho_{ratio,pop}}^{j}$ we use the median values given in Table \ref{tab:luque_post}.}
    \label{fig:luque_res_grid}
\end{figure}

We run this 27-planet sample through \texttt{RhoPop} with $N_c = 1$ and 2, i.e., 1- and 2-population models. We visualize our results in Figure \ref{fig:luque_res_grid} and summarize the hyperparameter posterior statistics in Table \ref{tab:luque_post}. We find the 2-population model is strongly preferred with a Bayes factor $B_{21} = 1042$. In frequentist parlance, this corresponds to a $p$-value of $3.4\times10^{-5}$ or a  4.1$\sigma$ detection of 2-populations. For good measure, we run this sample using the 3-population model. We find $B_{32} = 0.26$ meaning the 2-population model is indeed favored over the 3-population model.

As further validation, our 2-population model identifies the same 21 planets as \citet{Luque_2022} belonging to a higher density population. We find this population is best fit by a central WMF of 0.02. This WMF falls within the degenerate zone described in Sect. \ref{fig:comp_grid}. As such, we apply a $\chi^2$ minimization with these 21 planets over our full rock grid to find the equivalent CMF assuming the planets are water-free. We find a CMF = 0.28 (Fe/Mg = 0.67), meaning this population is either slightly wet or has a $\sim 6\%$ Fe-depletion relative to the average volatile-/water-free NRPZ composition of CMF = 0.29 (Fe/Mg = 0.71).

We find the same 6 planets that \citet{Luque_2022} identify as being 50\% water worlds as indeed belonging to a separate lower-density population. We, however, find this population is best fit by a central WMF of 0.78. While this WMF value is substantially more than that predicted by \citet{Luque_2022}, this is primarily due to the differences in the underlying material equations of state used in this work versus \citet{Luque_2022}. Both populations are found to have very little intrinsic scatter. Even at 10$M_\oplus$, where the difference in density ratios is minimized over the mass range considered, the two populations are separated by more than 12x their mutual intrinsic scatter parameters, meaning there is a robust density gap between these two small-planet populations.

\begin{table}[ht]
\large
\centering
\begin{tabular}{lll|ll|}
\cline{2-5}
\multicolumn{1}{l|}{}                                  & \multicolumn{2}{l|}{$N_c = 1$} & \multicolumn{2}{l|}{$N_c = 2$} \\ \hline
\multicolumn{1}{|l|}{Param}                            & Med            & 68\% CI            & Med            & 68\% CI         \\ \hline
\multicolumn{1}{|l|}{$\mu_{comp}^1$}               &  0.09 WMF            &  [0.14 WMF, 0.06 WMF] & 0.02 WMF  &  [0.03 WMF, 0.017 WMF]         \\
\multicolumn{1}{|l|}{$\sigma_{\rho_{ratio, pop}^1}$} &  0.21            &     [0.17, 0.25]         &   0.02            &  [0.005, 0.04]        \\
\multicolumn{1}{|l|}{$w^1$}                      &       1         &               &         0.76     &     [0.67, 0.83]      \\
\multicolumn{1}{|l|}{$\mu_{comp}^2$}               &                &               &    0.78 WMF       &  [0.87 WMF, 0.69 WMF]      \\
\multicolumn{1}{|l|}{$\sigma_{\rho_{ratio, pop}^2}$} &                &               &   0.03          &  [0.008, 0.07]        \\
\multicolumn{1}{|l|}{$w^2$}                      &                &               &     1-$w^1$          &              \\\hline
   
\multicolumn{1}{l|}{} & & $\ln \mathcal{Z}_1 = 7.20 \pm 0.24$ & & $\ln \mathcal{Z}_2 = 14.15 \pm 0.32$  \\\cline{2-5}

\end{tabular}

\vspace*{0.25 cm}

\begin{tabular}{|c|c|c|c|c|}
\hline
     $B_{21}$ & $p$-value & Sigma & Interpretation  \\\hline
     1042 & 3.4$\times 10^{-5}$ & 4.1$\sigma$ & \textit{Strong} evidence for two populations \\\hline
\end{tabular}

\caption{Summary of parameter posteriors, model evidence, and model selection parameters for our 1- and 2-population model runs with the \citet{Luque_2022} sample. CI = credible interval}.
\label{tab:luque_post}
\end{table}

\subsection{\citet{Adibekyan_2021} sample}
\label{sect:adib_results}
\begin{figure}[t]
    \centering
    \includegraphics[width = 0.99\textwidth]{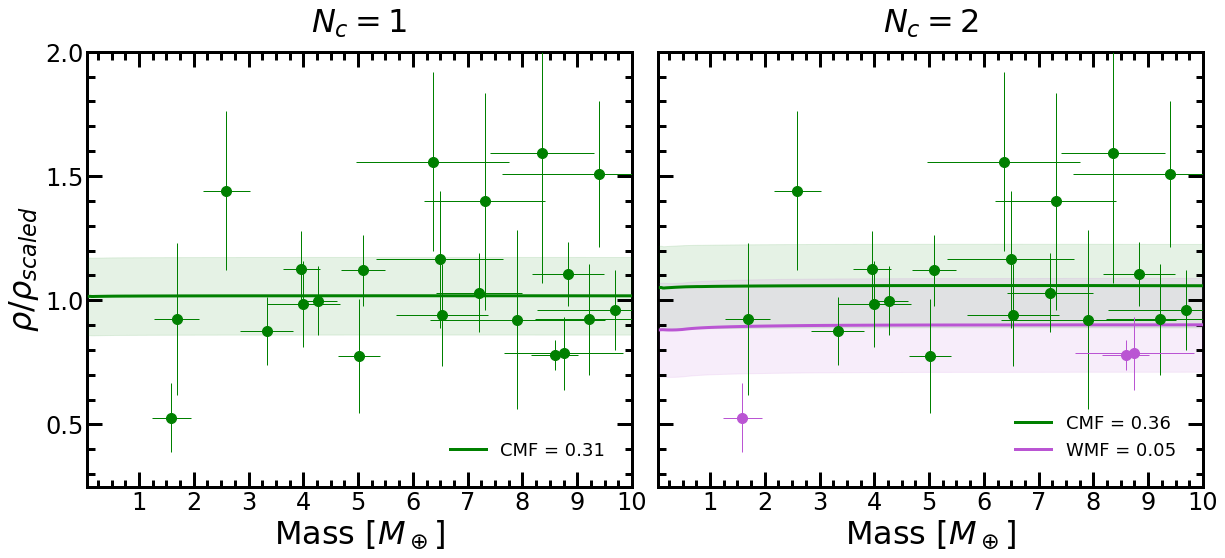}
    \caption{Density ratio as a function of mass for the preferred $N_c = 1$ model (left) and the $N_c = 2$ model (right)} applied to the \citet{Adibekyan_2021} sample. We color code the highest-density population as green and the lowest-density population as purple. The planets are assigned the same color as the population to which they most likely belong. The solid lines are the isocomposition lines for each $\mu_{comp}^{j}$. The bands are the corresponding intrinsic density ratio scatter parameter $\sigma_{\rho_{ratio,pop}}^{j}$. For all $\mu_{comp}^{j}$ and $\sigma_{\rho_{ratio,pop}}^{j}$ we use the median values given in Table \ref{tab:adibekyan_post}.
    \label{fig:adib_results_grid}
\end{figure}

The \citet{Adibekyan_2021} sample includes 22 likely volatile-free rocky planets with $M_p < 10 M_\oplus$ and mass and radius uncertainties $< 30\%$ around FGK stars. The authors identify a visually apparent, but not statistically significant, subpopulation of 5 iron-rich planets. We run our 1- and 2-population models with this sample and find $B_{21} = 0.39$, meaning there is no support for the more complex 2-population over the 1-population model. Thus, the single population model shown in Fig. \ref{fig:adib_results_grid} is our preferred model for this data set. Given the lack of support for the 2-population model for this data set, we do not consider the 3-population scenario. The results for the 2-population and preferred 1-population model are summarized in Table \ref{tab:adibekyan_post}.

\begin{table}[ht]
\large
\centering
\begin{tabular}{lll|ll|}
\cline{2-5}
\multicolumn{1}{l|}{}                                  & \multicolumn{2}{l|}{$N_c = 1$} & \multicolumn{2}{l|}{$N_c = 2$} \\ \hline
\multicolumn{1}{|l|}{Param}                            & Med            & 68\% CI    & Med & 68\% CI\\ \hline
\multicolumn{1}{|l|}{$\mu_{comp}^1$}               &  0.31 CMF      &  [0.02 WMF, 0.36 CMF] & 0.36 CMF & [0.30 CMF, 0.53 CMF]\\
\multicolumn{1}{|l|}{$\sigma_{\rho_{ratio, pop}^1}$} &  0.16          &     [0.11, 0.22]  & 0.17 & [0.06, 0.69]\\
\multicolumn{1}{|l|}{$w^1$}                      &       1        &     &    0.64 &   [0.07, 0.94]    \\
\multicolumn{1}{|l|}{$\mu_{comp}^2$}               &                &               &    0.05 WMF      &  [0.53 WMF, 0.33 CMF]      \\
\multicolumn{1}{|l|}{$\sigma_{\rho_{ratio, pop}^2}$} &                &               &   0.19         &  [0.09, 0.77]        \\
\multicolumn{1}{|l|}{$w^2$}                      &                &               &     1-$w^1$          &              \\\hline

\multicolumn{1}{l|}{} & & $\ln \mathcal{Z}_1 = -25.991 \pm 0.223$ & & $\ln \mathcal{Z}_2 = -26.93 \pm 0.21$  \\\cline{2-5}

\end{tabular}

\vspace*{0.25 cm}

\begin{tabular}{|c|c|c|c|c|}
\hline
     $B_{21}$ & $p$-value & Sigma & Interpretation  \\\hline
     0.39 & 1.0 & 0 & No evidence for more than one population \\\hline
\end{tabular}

\caption{Summary of parameter posteriors, model evidence, and model selection parameters for our 1- and 2-population model runs with the \citet{Adibekyan_2021} sample. CI = credible interval}.
\label{tab:adibekyan_post}
\end{table}

\section{Minimum number of planets required to resolve a population of iron-rich planets}
\label{sect:n_planet_estimates}
Here we assume the 5 potentially iron-rich planets identified by \citet{Adibekyan_2021} do indeed represent a distinct higher-density population from a lower-density population comprised of the remaining 17 planets and that our null finding in Sect. \ref{sect:adib_results} is a result of observational uncertainty alone. We then estimate the minimum number of planets needed to resolve a population of iron-enriched planets from a lower-density population of volatile-/water-free rocky planets. Evidencing the Solar System scaled-density dichotomy between Mercury and the other rocky planets, we assume the lower-density population has an Earth-like composition, i.e., $\mu_{comp}^\mathrm{el} = 0.325$ CMF, and the higher density population has a Mercury-like core mass fraction of $\mu_{comp}^\mathrm{ml} = 0.7$ CMF. The latter choice is further supported by the fact that $\chi^2$ is minimized between the 5 Fe-rich planets of \citet{Adibekyan_2021} and our full rock grid when CMF$\sim$ 0.73. For a sense of scale, the Hypatia catalog has Fe/Mg = $0.79 \pm 0.18$ corresponding to CMF =  $0.29 \pm 0.05$, meaning this population is $\sim 8\sigma$ above the central CMF of Hypatia. Per the \citet{Adibekyan_2021} sample, Fe-rich planets represent 0.23 of the total population. For simplicity, we round this up to a quarter and set the mixing weight of our Mercury-like population to $w^\mathrm{ml} = 0.25$. The corresponding mixing weight for the Earth-like population is then $w^\mathrm{el} = 0.75$.

For each sample, we randomly draw $N_p$ planet masses from $M_p \sim \mathcal{U}[1, 10] M_\oplus$. We then assume $N_{ml} = w_{ml}\times N_p$ of these are Mercury-like and $N_{el} = w_{el}\times N_p$ are Earth-like in composition. We then calculate the density corresponding to $\mu_{comp}^\mathrm{ml} = 0.7$ CMF and $\mu_{comp}^\mathrm{el} = 0.325$ CMF at mass for each Mercury-like and Earth-like planet, respectively, and then calculate the corresponding planet radius, $R_p$. Finally, we resample each mass and radius from the Gaussian distributions $\sim\mathcal{N}(M_p, \sigma_{M_p})$ and $\sim\mathcal{N}(R_p, \sigma_{R_p})$, respectively, to simulate observational uncertainties. We consider three sets of mass-radius uncertainties: (1) $\sigma_{M_p} = 10\%$ and $\sigma_{R_p} = 2.5\%$, (2) $\sigma_{M_p} = 6\%$ and $\sigma_{R_p} = 2\%$, and (3) $\sigma_{M_p} = 5\%$ and $\sigma_{R_p} = 1\%$. For each set of mass-radius uncertainties, we calculate $B_{21}$ and $n_\sigma$ as a function of $N_p$.

We find that the number of planets required to achieve a strong detection of both populations is highly dependent on the average mass and radius uncertainties. For the near best-case scenario of $\sigma_{M_p} = 5\%$ and $\sigma_{R_p} = 1\%$, we estimate only $\sim 12$ planets are needed to detect both populations with moderate support ($B_{21} > 12$ and $n_\sigma > 2.7$) and $\sim 16$ for detection with strong support ($B_{21} > 150$ and $n_\sigma > 3.6$). For the most conservative, but still precise, case of $\sigma_{M_p} = 10\%$ and $\sigma_{R_p} = 2.5\%$, a moderate detection requires approximately 117-118 planets, and a strong detection requires at least $\sim$ 154 planets, i.e., an almost factor of 10 increase in $N_p$ from the near best-case scenario. For comparison, the average mass and radius uncertainties for the \citet{Adibekyan_2021} sample are $14\%$ and $5\%$, respectively. As such, it is not surprising that they do not find a conclusive and distinct population of Fe-rich planets as it would require significantly more than 154 planets to do so at their average precisions. We also show our model results for the synthetic 20-planet sample with $\sigma_{M_p} = 5\%$ and $\sigma_{R_p} = 1\%$ in detail in Appendix \ref{sect:toy_sample_example} to demonstrate \texttt{RhoPop}'s ability to recover known hyperparameters.

\begin{figure}[ht]
    \centering
    \includegraphics[width = 0.5\textwidth]{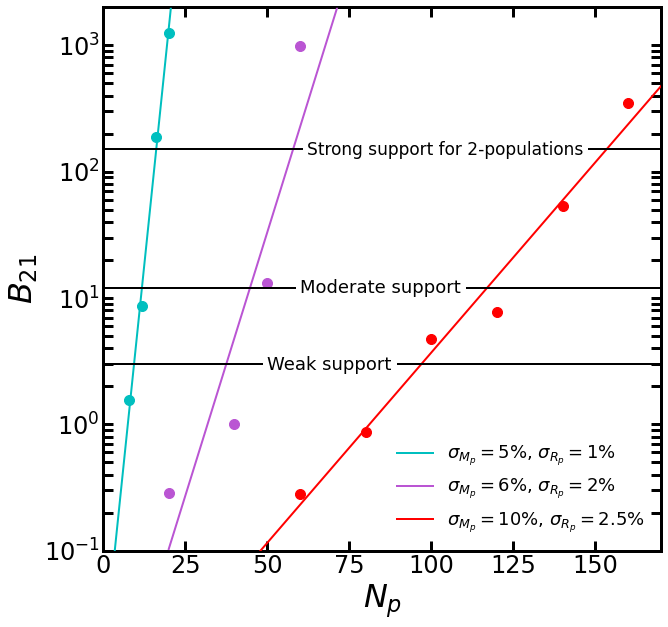}
    \caption{$B_{21}$ (left vertical axis) and $n_\sigma$ (right vertical axis) as a function of $N_p$ for average observational uncertainties of $\sigma_{M_p} = 5\%$ and $\sigma_{R_p} = 1\%$ (cyan), $\sigma_{M_p} = 6\%$ and $\sigma_{R_p} = 2\%$ (purple), and $\sigma_{M_p} = 10\%$ and $\sigma_{R_p} = 2.5\%$ (red). }
    \label{fig:enter-label}
\end{figure}

\section{Discussion}

\subsection{Comparisons with the nominally rocky planet zone}
 We compare our results for the \citet{Luque_2022} (Sect. \ref{sect:Luque_results}) and \citet{Adibekyan_2021} (Sect. \ref{sect:adib_results}) samples with the nominally rocky planet zone \citep{Unterborn_2022}. From Eq. 24 and 27 of \citet{Unterborn_2022}, we estimate that the NRPZ covers a density ratio range of 0.85-1.1. We reiterate that the NRPZ represents the expected 3$\sigma$ compositional range of volatile-free and water-free rocky planets. The preferred single population model for the \citet{Adibekyan_2021} sample yields a 1$\sigma$ density ratio range that is similar in width to the NRPZ, meaning the 3$\sigma$ range for this sample is $\sim 3\times$ larger than the NRPZ. This result is not surprising given the number of exceptions to this simplistic picture of planet formation we outline in Sect. \ref{sect:intro}.

Our 1-population model for the \citet{Luque_2022} sample finds a central composition of 0.09 WMF with an intrinsic 1$\sigma$ scatter of $\sigma_{\rho_{pop}} = 0.21$. This leads to a 1$\sigma$ density range of $\sim 0.6-1.0$ which is \textit{inconsistent} with the NRPZ. This result is not surprising since we expected to find 2 populations a priori given the results of \citet{Luque_2022}, although we did not inject this expectation into our models. This tension with the NRPZ is reduced for the strongly preferred 2-population model. We find a higher density population with a central composition of WMF = 0.02 and $\sigma_{\rho_{pop}} = 0.02$. The 2\% water mass fraction isocomposition line corresponds to a density ratio of $\sim 0.95$ and the 3$\sigma$ range in density ratios is $\sim 0.89-1.01$. This density ratio range falls entirely within the NRPZ meaning that this population of planets can be explained via volatile-free rocky compositions, i.e., we don't need to invoke water to explain the densities of these planets. More notable, however, is the fact that the 3$\sigma$ range of this population is approximately half the width of the NRPZ suggesting the compositions of these planets may be less diverse than stellar abundances predict. We note the important caveat that the NRPZ is estimated using Fe, Mg, and Si abundances from the Hypatia Catalog database \citep{Hinkel_2014} and is thus strongly biased towards FGK stars as there is a lack of M-type stars with abundance measurements, especially those with Fe, Mg, \textit{and} Si measurements. This suggests that the \citet{Luque_2022} sample has a smaller spread in the major rock-building elements than FGK stars. In fact, M dwarfs that host planets in general may have a smaller spread. Testing this possibility, however, requires more M-dwarf abundances of Fe, Mg, and Si to build an M-dwarf NRPZ, i.e., an M-NRPZ. While TOI-561 b in the \citet{Adibekyan_2021} sample has a $\rho_{ratio} \sim 0.5$, consistent with the lower-density population of \citet{Luque_2022}, the lack of evidence for a water-rich population around FGK stars further suggests that small-planet compositions and formation processes may be dependent on host-star type.

\subsection{The TRAPPIST-1 planets}
The primary goal of Sect. \ref{sect:Luque_results} is to validate the results of \citet{Luque_2022} using their sample as published. We note, however, that the TRAPPIST-1 system accounts for roughly a quarter of the LP22 sample. Further, the seven TRAPPIST-1 planets have some of the smallest error bars with an average density uncertainty of 5.6\% compared to the remainder of the sample which has an average density uncertainty of 19\%. Together this suggests our detection of 2-populations of M dwarf planets may be driven in large part by the TRAPPIST-1 system. To test this, we removed the TRAPPIST-1 planets from the LP22 sample and reran the 1- and 2-population models. Without these seven planets, the detection of 2-populations of planets is reduced from 4.1$\sigma$ to 1.8$\sigma$, meaning the TRAPPIST-1 planets do indeed play a large role in the detection of 2-populations of small planets around M dwarfs. Further, for the $N_c = 2$ scenario, we find a median value of $\mu^1_{comp} = 0.327$ CMF, nearly identical to the CMF of Earth, in contrast to $\mu^{1}_{comp} = 0.02$ WMF, when the TRAPPIST-1 planets are included. This begs the question as to whether volatile-/water-free rocky planets around M dwarfs are generally Earth-like in composition and the TRAPPIST-1 planets are anomalously iron-depleted or wet. A larger sample of small M dwarf planets with density precisions comparable to the TRAPPIST-1 planets and/or an M-NRPZ are needed to fully investigate this question.

\subsection{Alternative explanations for water worlds}
We note that our water grid assumes condensed water/ice phases. The equilibrium temperatures of the five LP22 planets in the lower density population range from $\sim 1.5 - 4.2\times$ Earth's. As such, it is likely substantial fractions of their water exist in the vapor and super-critical phases. Our choice of condensed water/ice then overestimates the density of these water layers, and, therefore, our WMF estimates serve as an upper limit. We chose to work in terms of condensed water/ice for the most direct comparison with the interpretations made in \citet{Luque_2022}, but a future iteration of \texttt{RhoPop} will incorporate the super-critical and vapor phases of water and planet equilibrium temperature.

We also note that the population of planets interpreted as water-worlds in \citet{Luque_2022} and this work is not uniquely explained as such. \citet{Rogers_james_2023} show that these planets are equally well explained as sub-Neptunes that have retained a small fraction of their primordial H/He envelopes and provide some observational evidence to support this interpretation. It is also plausible such planets have outer gaseous envelopes comprised of species heavier than H/He \citep[e.g.,][]{Piaulet_2023}. In short, the source M dwarf planet density gap remains to be fully explored. In a future update, we will expand \texttt{RhoPop} to account for these various scenarios and investigate whether the data favors one scenario over the others.

\subsection{Caveats to estimates on the minimum number of planets needed to resolve a Fe-rich population}
In Sect. \ref{sect:n_planet_estimates}, we estimated that approximately 154 planets with average $\sigma_{M_p} = 10\%$ and $\sigma_{R_p} = 2.5\%$ are needed to resolve a population of planets with Mercury-like compositions from a population with nominally Earth-like compositions. Interestingly, there are 154 small planets on the NASA Exoplanet Archive\footnote{as of 07 AUG 2023} with density uncertainties less than 100\%. The average and median M-R uncertainties of these planets, however, are $\sigma_{M_p} = 27\%$ and $\sigma_{R_p} = 6\%$ and $\sigma_{M_p} = 19\%$ and $\sigma_{R_p} = 5\%$, respectively. These uncertainties are $\gtrsim 2\times$ the precisions required to resolve a Fe-rich population with high confidence for $N_p = 154$. As such, even if such a population exists, we do not expect it to be resolvable with the current sample of small planets. We note, however, that our analysis assumes there is 1 Fe-rich planet for every 3 nominally Earth-like planets ($w^{ml} = 0.25$) based on the work of \citet{Adibekyan_2021}. If Fe-rich planets are more abundant relative to Earth-like planets than 1:3, then the required sample size to resolve this population will be reduced and vice versa. Further, we assume a constant bulk composition for the iron-rich population of $\mu_{comp}^{ml} = 0.7$. While we use Mercury and the 5 potentially Fe-rich planets from \citet{Adibekyan_2021} to justify this choice, the average CMF can be higher or lower which will act to make this population more or less easily resolvable at a given mass and radius precision, respectively. We will consider a wider range of mixing weights and compositions for the Fe-rich population in future work.

\section{Conclusions}
We present the open-source software \texttt{RhoPop} for identifying distinct populations of small planets and use it to validate the recent results of \citet{Luque_2022} and \citet{Adibekyan_2021}. While we show that a distinct population of Fe-rich planets is likely unresolvable with the current sample of small planets, \texttt{RhoPop} is designed to provide the community with a ready-to-go tool for identifying such populations as the number and precisions of small planets continue to increase. The next release of \texttt{RhoPop} will include pre-loaded compositional grids with atmosphere layers comprised of various molecular species. We will use these additional grids to investigate whether the population of planets interpreted by \citet{Luque_2022} as 50\% water worlds are better explained as having an atmosphere. In future work, we will apply \texttt{RhoPop} to the entire available sample of small planets and expand our analysis on the minimum number of planets required to resolve a population of iron-rich planets.

\section{Acknowledgements}
W.R.P's effort is based upon work while serving at the National Science Foundation and was initially funded by NSF grant EAR-1724693. Work by B.S.G. was supported by the Thomas Jefferson Chair for Space Exploration endowment from The Ohio State University. C.T.U acknowledges support under grants NNX15AD53G and 80NSSC23K0267 from the National Aeronautics and Space Administration (NASA). The results reported herein benefited from collaborations and/or information exchange within NASA's Nexus for Exoplanet System Science (NExSS) research coordination network sponsored by NASA's Science Mission Directorate. This work is supported by the National Science Foundation under Grant No. 2143400.

\newpage

\begingroup
\renewcommand{\section}[2]{}%

\bibliography{biblio.bib}
\endgroup

\newpage
\section*{Appendix}
\section{Synthetic planet sample example}
\label{sect:toy_sample_example}

Here we show the detailed results of \texttt{RhoPop} applied to a synthetic 20-planet sample where the hyperparameters are known a priori to demonstrate their successful recovery. To reiterate, we sample 5 planets from the Mercury-like population with $\mu_{comp}^{ml} = 0.7$ CMF and intrinsic scatter $\sigma_{\rho_{ratio, pop}^1}$ = 0.0. Similarly, we sample 15 planets from an Earth-like population with $\mu_{comp}^{el} = 0.325$ CMF and intrinsic scatter $\sigma_{\rho_{ratio, pop}^2}$ = 0.0. This gives $w^{el} = 0.75$ and $w^{ml} = 1-w^{el} = 0.25$. We use the same hyperparameter priors given in Table \ref{tab:hp_priors} and assign 5\% uncertainties in $M_\mathrm{ob}$ and 1\% in $R_\mathrm{ob}$ to each planet.

\begin{figure}[ht]
    \centering
    \includegraphics[width = 0.9\textwidth]{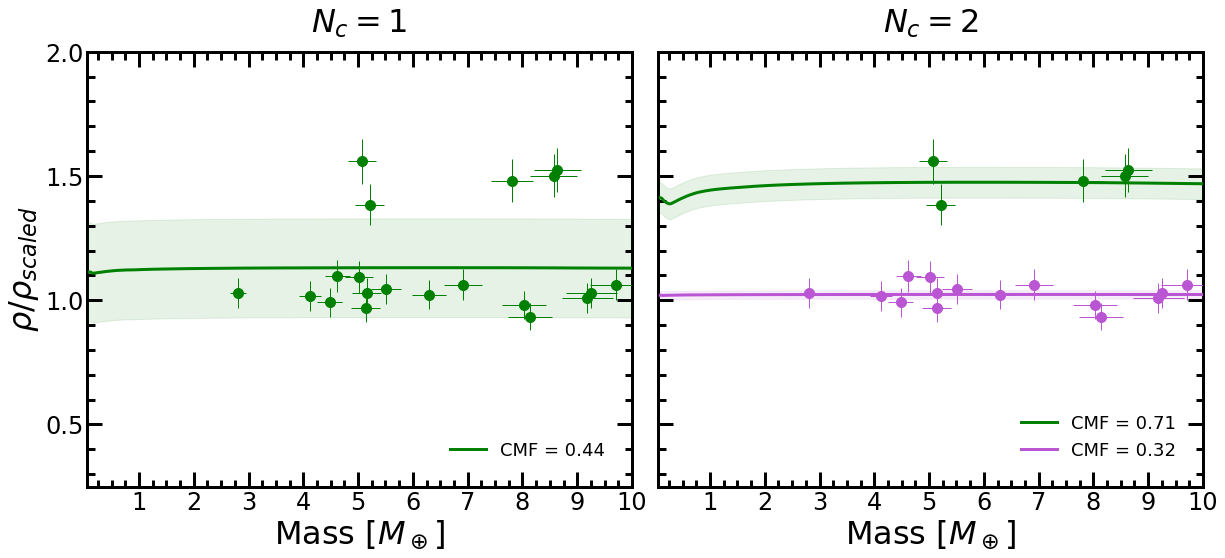}
    \caption{Summary of results for a synthetic 20-planet sample of planets with two-populations: a Mercury-like one with $\mu_{comp}^{ml} = 0.7$ CMF and $w^{ml} = 0.25$ and an Earth-like population with $\mu^{el}_{comp} = 0.325$ CMF and $w^{el} = 0.75$.}
    \label{fig:20_planet_res_grid}
\end{figure}

\begin{figure}[ht]
    \centering
    \includegraphics[width = 0.5\textwidth]{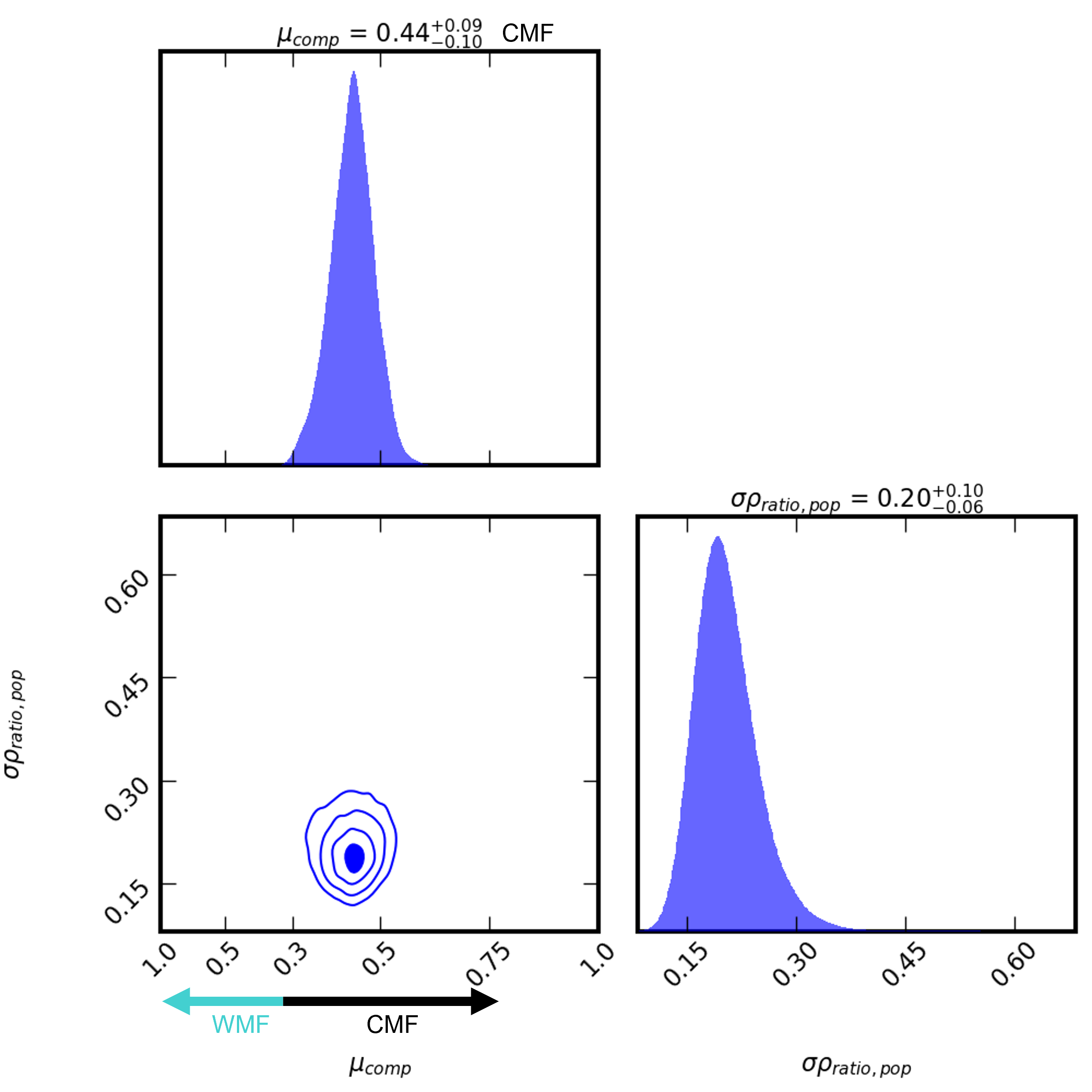}
    \caption{hyperparameter posterior corner plot for the 1-population model applied to our synthetic sample. The +/- values correspond to the 95\% credible} intervals. This figure was generated with the \texttt{dynesty} dynamic nested sampling software package \citep{speagle20_dynesty}.
    \label{fig:20planet_sample_1pop_corner}
\end{figure}

\begin{figure}[ht]
    \centering
    \includegraphics[width = 0.75\textwidth]{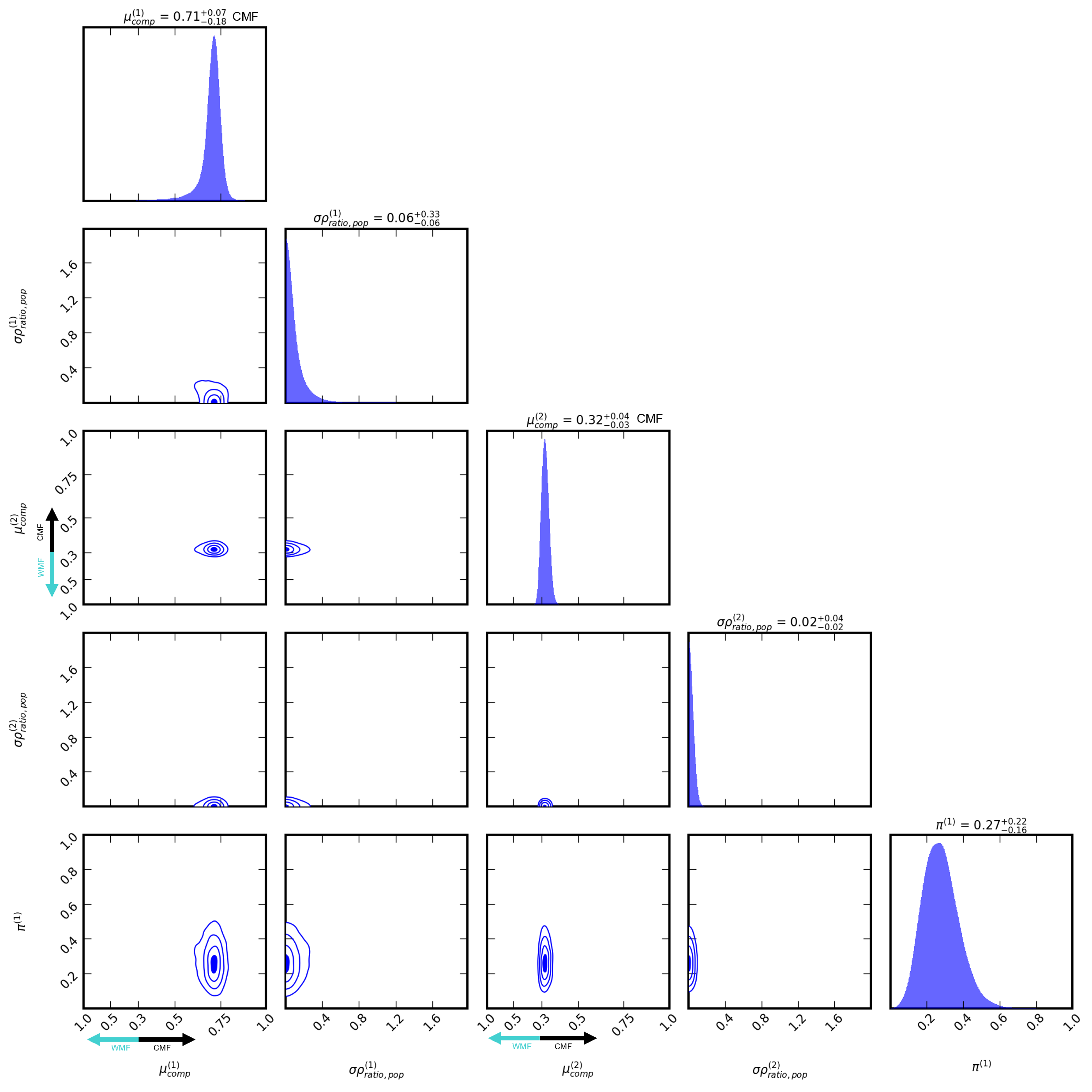}
    \caption{hyperparameter posterior corner plot for the 2-population model applied to our synthetic sample. The +/- values correspond to the 95\% credible} intervals. This figure was generated with the \texttt{dynesty} dynamic nested sampling software package \citep{speagle20_dynesty}.
    \label{fig:20planet_sample_2pop_corner}
\end{figure}

We apply the 1- and 2-population \texttt{RhoPop} models to this sample and find, unsurprisingly, that the 2-population model is favored over the 1-population model at the $> 4\sigma$ level. The results are summarized in Table \ref{tab:20planet_sample} and visualized in Fig. \ref{fig:20_planet_res_grid}. The 1-population model is naturally a poor fit for this sample and recovers a mean composition of CMF = 0.42 which is $\simeq w^{el}\times \mu_{comp}^{el} + w^{ml}\times \mu_{comp}^{ml}$.  None of the true hyperparameter values are contained within the 68\% CIs of the estimated $\mu_{comp}$ for the 1-population model. The 2-population model, however, recovers accurate values for both the Earth-like and Mercury-like populations. \texttt{RhoPop} finds $\mu_{comp}^{1} = 0.71^{+0.03}_{-0.04}$ CMF corresponding to the Mercury-like population we fixed to have $\mu_{comp}^{ml} = 0.7$ CMF. Similarly, \texttt{RhoPop} finds a lower density population with $\mu_{comp}^{2} = 0.32^{+0.02}_{-0.02}$ CMF corresponding to the Earth-like population we fixed to have $\mu_{comp}^{el} = 0.325$ CMF. Further, this model finds $w^{1} = 0.27^{-0.09}_{+0.1}$ which is consistent with the injected value of $w^{ml} = 0.25$. We show the hyperparameter posteriors for the 1- and 2-population models in Fig. \ref{fig:20planet_sample_1pop_corner} and Fig. \ref{fig:20planet_sample_2pop_corner}, respectively.

\begin{table}[ht]
\large
\centering
\begin{tabular}{lll|ll|}
\cline{2-5}
\multicolumn{1}{l|}{}                                  & \multicolumn{2}{l|}{$N_c = 1$} & \multicolumn{2}{l|}{$N_c = 2$} \\ \hline
\multicolumn{1}{|l|}{Param}                            & Med            & 68\% CI            & Med            & 68\% CI         \\ \hline
\multicolumn{1}{|l|}{$\mu_{comp}^1$}               &  0.44 CMF            &  [0.39 CMF, 0.48 CMF] & 0.71 CMF  &  [0.67 CMF, 0.74 CMF]       \\
\multicolumn{1}{|l|}{$\sigma_{\rho_{ratio, pop}^1}$} &  0.20           &     [0.16, 0.24]         &   0.06            &  [0.02, 0.17]        \\
\multicolumn{1}{|l|}{$w^1$}                      &       1         &               &         0.27     &     [0.18, 0.37]      \\
\multicolumn{1}{|l|}{$\mu_{comp}^2$}               &                &               &    0.32 CMF       &  [0.30 CMF, 0.34 CMF]      \\
\multicolumn{1}{|l|}{$\sigma_{\rho_{ratio, pop}^2}$} &                &               &   0.02          &  [0.005, 0.04]        \\
\multicolumn{1}{|l|}{$w^2$}                      &                &               &     1-$w^1$          &              \\\hline
   
\multicolumn{1}{l|}{} & & $\ln \mathcal{Z}_1 = -3.34 \pm 0.20$ & & $\ln \mathcal{Z}_2 = 3.79 \pm 0.27$  \\\cline{2-5}

\end{tabular}

\vspace*{0.25 cm}

\begin{tabular}{|c|c|c|c|c|}
\hline
     $B_{21}$ & $p$-value & Sigma & Interpretation  \\\hline
     1250 & 2.8$\times 10^{-5}$ & 4.2$\sigma$ & \textit{Strong} evidence for two populations \\\hline
\end{tabular}

\caption{Summary of parameter posteriors, model evidence, and model selection parameters for our 1- and 2-population model for a synthetic sample of 20 planets with Mercury-like and Earth-like subpopulations. CI = credible interval}.
\label{tab:20planet_sample}
\end{table}

\end{CJK*}

\end{document}